\documentclass[a4paper,12pt]{amsart}
\usepackage{amssymb,amsthm}
\usepackage{amsmath,amstext}
\usepackage{amscd}
\usepackage{mathrsfs}
\usepackage{bbm}
\usepackage{bbold}
\usepackage{graphicx}
\usepackage{psfrag}
\usepackage{multirow}
\usepackage{hyperref}
\usepackage[usenames,dvipsnames,svgnames,table]{xcolor}
\usepackage{rotating}
\usepackage{ifthen}
\usepackage[headings]{fullpage}

\usepackage{tikz}
\usetikzlibrary{matrix,positioning}
\usetikzlibrary{arrows}
\usetikzlibrary{decorations.markings}
\usetikzlibrary{shapes.geometric}
\usetikzlibrary{shapes.misc}
\usetikzlibrary{calc}

\newtheorem{theorem}{Theorem}[section]

\theoremstyle{definition}

\newtheorem{example}[theorem]{Example}

\theoremstyle{remark}

\numberwithin{equation}{section}

\newcommand{\ds}{\displaystyle}

\newcommand{\1}{\bar{1}}
\newcommand{\catalan}[1]{C_{#1}}
\newcommand{\ballot}[2]{C^{#1}_{#2}}
\DeclareMathOperator{\rate}{rate}
\newcommand{\dstar}{$D^*$-TASEP}
\newcommand\bra[1]{{\langle#1|}}
\newcommand\ket[1]{{|#1\rangle}}

\begin{document}

\title
[The exact phase diagram for the $D^*$-TASEP]
{The exact phase diagram for a semipermeable TASEP
with nonlocal boundary jumps}

\author{Erik Aas}
\address{Erik Aas, Department of mathematics, Pennsylvania State University, PA 16802, USA}
\email{eaas@kth.se}

\author{Arvind Ayyer}
\address{Arvind Ayyer, Department of mathematics, Indian Institute of Science, Bangalore - 560012, India}
\email{arvind@iisc.ac.in}

\author{Svante Linusson}
\address{Svante Linusson, Department of mathematics, KTH, SE-10044 Stockholm, Sweden}
\email{linusson@math.kth.se}

\author{Samu Potka}
\address{Samu Potka, Department of mathematics, KTH, SE-10044 Stockholm, Sweden}
\email{potka@kth.se}

\date{\today}

\begin{abstract}
We consider a finite one-dimensional totally asymmetric simple exclusion process (TASEP) with four types of particles, $\{1,0,\bar{1},*\}$, in contact with reservoirs. Particles of species $0$ can neither enter nor exit the lattice, and those of species $*$ are constrained to lie at the first and last site. Particles of species $1$ enter from the left reservoir into either the first or second site, move rightwards, and leave from either the last or penultimate site. Conversely, particles of species $\bar{1}$ enter from the right reservoir into either the last or penultimate site, move leftwards, and leave from either the first or last site.
This dynamics is motivated by a natural random walk on the Weyl group of type D.
We compute the exact nonequilibrium steady state distribution using a matrix ansatz building on earlier work of Arita. We then give explicit formulas for the nonequilibrium partition function as well as densities and currents of all species in the steady state, and derive the phase diagram.
\end{abstract}

\subjclass[2010]{82C22, 82C23, 82C26, 60J27}
\keywords{exclusion process, semipermeable, two-species, nonlocal boundary jumps, phase diagram, current, fat shock}

\maketitle

\section{Introduction}
\label{sec:intro}

Exact solutions of nonequilibrium statistical mechanical models have served as useful testbeds for potential theories of nonequilibrium 
statistical mechanics. The most prominent example of this phenomenon in recent times is the totally asymmetric simple exclusion process (TASEP) on a finite one-dimensional lattice, whose steady state was obtained by Derrida, Evans, Hakim and Pasquier~\cite{dehp}. The exact solution via the matrix ansatz also enabled the computation of the large deviations of density~\cite{derrida-lebowitz-speer-1997} and the current~\cite{lazarescu-mallick-2011}. These computations were the first nontrivial test of two key ideas: the {\em additivity principle}~\cite{bodineau-derrida-2004} and the {\em macroscopic fluctuation theory}~\cite{BDJGL-2005}.

Exclusion processes with multiple species of particles are natural candidates for the modelling of systems where there are several subpopulations in a large population~\cite{simpson-et-al-2009,bruna-chapman-2012}. 
Prominent examples of these include movements of (human or animal) crowds, traffic flow and heterogeneous ion transport.

The steady state of the TASEP with two species (and vacancies) with periodic boundaries was determined exactly by the matrix ansatz~\cite{djls}, but the two-species process with open boundaries and arbitrary rates is not believed to have a simple solution. With special kinds of boundary rates, the steady states have been determined exactly using the matrix ansatz in several cases~\cite{evans-et-al-1995, arita-2006, als-2009, als-2012, CMRV-2015, CEMRV-2016, afr-2018}.

In this work, we consider a variant of the semipermeable TASEP~\cite{arita-2006, als-2009}. Recall that this is a TASEP with two species of particles: $1$ (first-class particles), $2$  (second-class particles) and vacancies labelled $0$, where both $1$'s and $2$'s hop right, and $1$'s preferentially hop over the $2$'s. First-class particles enter from the left with rate $\alpha$ and exit to the right with rate $\beta$. The semipermeability comes from the fact that $2$'s cannot leave the system.

In our model, we have particles labelled $\{1,0,\bar{1},*\}$ on a one-dimensional lattice of length $n$. We dub this the {\em \dstar{}} because of its relation with Lam's random walk~\cite{lam-2015} for the affine Weyl group of type D.
For more on this, see Section~\ref{sec:discuss}.
It will be convenient for us to think of particles $1$ and $\bar 1$ being charged positively and negatively respectively. Particles of type $0$ are vacancies, and those of type $*$ are neutral. However, there are some important differences. 
Each site in positions $2,\dots,n-1$ can be occupied by one of $\bar 1, 0$ and $1$. The sites at the boundaries can only be occupied by either $0$ or $*$. 
The dynamics of the \dstar{} can be thought of as the action of a rightward-pointing electric field, and is as follows. 
In the bulk,
\begin{equation}
\label{bulkrates}
\begin{split}
1\bar{1} & \to \bar{1}1 \quad \text{with rate $1$}, \\
10 & \to 01 \quad \text{with rate $1$}, \\
0\bar{1} &\to \bar{1}0 \quad \text{with rate $1$}.
\end{split}
\end{equation}
At the left boundary, the transitions are 
\begin{equation}
\label{leftrates}
\begin{split}
*\bar{1} & \to *1 \quad \text{with rate $\alpha$}, \\
*0 & \to 01 \quad \text{with rate $\alpha_*$}, \\
0\bar{1} & \to *0 \quad \text{with rate 1}. 
\end{split}
\end{equation}
Finally, at the right boundary, the transitions are 
\begin{equation}
\label{rightrates}
\begin{split}
1* &\to \bar{1}* \quad \text{with rate $\beta$}, \\
0* &\to \bar{1}0 \quad \text{with rate $\beta_*$},  \\
10 &\to 0* \quad \text{with rate $1$} 
\end{split}
\end{equation}
At the left boundary, either a $1$ enters the second site, or a $\bar{1}$ leaves from the second site, or both, and conversely at the right boundary. This is consistent with the action of the electric field.
It is clear that the total number of $0$'s is fixed by the dynamics and we set it to be $n_0$. 
The \dstar{} exhibits {\em charge-conjugation symmetry}: if we simultaneously interchange $1$'s and $\bar{1}$'s, $\alpha$ and $\beta$, 
$\alpha_*$ and $\beta_*$, as well as  left and right directions, the \dstar{} is invariant. This fact will be useful in the analysis.

Let the state space be given by 
\begin{equation}
\Omega_{n,n_0} = \left\{ \tau \in \{\1,0,1,*\}^n \; \left| \; 
\substack{\ds n_0(\tau) = n_0, \; \tau_1, \tau_n \in \{0,* \}, \\[0.15cm]
\ds \tau_i \in \{\1,0,1\} \text{ for } 2 \leq i \leq n-1 } \right. 
\right\},
\end{equation}
where $n_0(\tau)$ is the number of $0$'s in $\tau$.
A little thought shows that if $\alpha,\alpha_*,\beta,\beta_*>0$ the \dstar{} is ergodic, i.e., it is possible to get from any configuration in $\Omega_{n,n_0}$ to any other. 
If any of the parameters is zero the model will not be ergodic. 
For example, in the case when $\alpha_* = \beta_* = 0$, 
 the first and last sites will eventually be occupied by $*$, and the resulting model is equivalent to the semipermeable TASEP on sites $2,\dots,n-1$.

The plan of the article is as follows. In Section~\ref{sec:matrix}, we show that the matrix ansatz can be used to construct the steady state probabilities. We also construct an explicit representation. In Section~\ref{sec:partfn}, we give an exact formula for the nonequilibrium partition function. We compute the current and densities exactly in Section~\ref{sec:curr-dens}. Using these results we derive the phase diagram in the thermodynamic limit in Section~\ref{sec:phase}. 
The computations of the thermodynamic limit are not mathematically rigorous, but we believe the resulting limits are correct. 
Our computations follow the same strategy as that of Arita~\cite[Section~5]{arita-2006}. 
We have also performed extensive numerical checks to convince ourselves that the result is true.  
The asymptotic computations that we need for the phase diagram are relegated to Appendix~\ref{sec:asymp}. We end with some discussion on the relationship of the \dstar{} to other TASEPs in Section~\ref{sec:discuss}.

\section{Matrix Ansatz}
\label{sec:matrix}

We will now determine the steady state using the matrix ansatz.
Since the \dstar{} is ergodic, the steady state is uniquely determined. It therefore suffices to verify the master equation. 
This is done using the matrix ansatz~\cite{dehp}.

For each species $j$, we will associate a matrix $X_j$. The steady state probability $\pi(\tau)$ of a configuration 
$\tau \in \Omega_{n,n_0}$ is then given by
\begin{equation}
\label{statprob}
\pi(\tau) = \frac{w(\tau)}{Z_{n,n_0}},
\end{equation}
where the {\em weight} of $\tau$ is given by
\begin{equation}
w(\tau) = \bra W X_{\tau_1} \cdots X_{\tau_n} \ket V,
\end{equation} 
and the nonequilibrium partition function $Z_{n,n_0}$ is then given by
\begin{equation}
\label{pf-dstar-matrix ansatz}
Z_{n,n_0} = \sum_{\tau \in \Omega_{n,n_0}} 
\bra W X_{\tau_1} \cdots X_{\tau_n} \ket V.
\end{equation}
We claim that the matrices $X_+,X_-,X_0,X_*$ and boundary vectors $\bra W, \ket V$ satisfy a particular algebra, which we now state. 
The bulk relations are the same as the usual two-species TASEP on a ring~\cite{djls} and the  semipermeable TASEP~\cite{arita-2006, als-2009}, namely
\begin{equation}
\label{bulkansatz}
\begin{split}
X_+ X_- &= X_+ + X_-, \\
X_+ X_0 &= X_0, \\
X_0 X_- &= X_0.
\end{split}
\end{equation}
The boundary relations are more complicated than that of the semipermeable TASEP.
On the left boundary, they are given by
\begin{equation}
\label{leftansatz}
\begin{split}
\bra W X_* X_0 &= \frac{1}{\alpha_*} \bra W X_0, \\
\bra W X_* X_- &= \frac{1}{\alpha} \bra W X_*. 
\end{split}
\end{equation}
On the right boundary, the relations are similarly
\begin{equation}
\label{rightansatz}
\begin{split}
X_0 X_* \ket V &= \frac{1}{\beta_*} X_0 \ket V, \\
X_+ X_* \ket V &= \frac{1}{\beta} X_* \ket V.
\end{split}
\end{equation}

The proof that the matrix ansatz equations \eqref{bulkansatz}\eqref{leftansatz} and \eqref{rightansatz} give the correct stationary probabilities using \eqref{statprob} are now standard and can be seen in \cite[Section 3.1]{blythe-evans-2007}. Both the domain-based proof and the algebraic proof work exactly the same way as for the semipermeable TASEP~\cite{arita-2006}. What is new here is that the boundary relations are also quadratic, but this presents no additional difficulty, as we now show.

Recall that the master equation for configuration $\tau$ can be written as
\begin{equation}
\label{mastereq}
\sum_{\tau' \in \Omega_{n,n_0}} \left( \rate(\tau\to \tau') \;
\pi(\tau) - \rate(\tau' \to \tau) \; \pi(\tau') \right) = 0.
\end{equation}
We rewrite \eqref{mastereq} as $\sum _{i=1}^{n-1} T_i (\tau) = 0$,
where $T_i$ only considers transitions involving sites $i$ and $i+1$. In other words,
\begin{equation}
\label{Ti-def}
T_i(\tau) = \rate(\tau_i\tau_{i+1} \to \tau_{i+1} \tau_i) \; \pi(\dots, \tau_i, \tau_{i+1}, \dots )      
- \rate(\tau_{i+1}\tau_i \to \tau_i\tau_{i+1}) \;
\pi(\dots, \tau_{i+1}, \tau_i, \dots),
\end{equation}
for $2 \leq i \leq n-2$.
Similarly,  $T_1(\tau), T_{n-1}(\tau)$ are defined by the boundary transitions \eqref{leftrates} and \eqref{rightrates} respectively. 

We now claim that the matrix algebra in  \eqref{bulkansatz}, \eqref{leftansatz} and \eqref{rightansatz}
allows us to show that 
\begin{equation}
\label{Ti-equations}
\begin{split}
  T_1 &= -a_{\tau_2} \; \pi(\tau_1, \hat{\tau}_2, \dots, \tau_n), \\
  T_i &= a_{\tau_i} \; \pi(\tau_1, \dots, \hat{\tau}_i, \dots, \tau_n) - 
  a_{\tau_{i+1}} \; \pi(\tau_1, \dots, \hat{\tau}_{i+1}, \dots, \tau_n ) \quad
  \text{for $2 \leq i \leq n-2$}, \\
  T_{n-1} &= a_{\tau_{n-1}} \; 
  \pi(\tau_1, \dots, \hat{\tau}_{n-1}, \tau_n),
\end{split}
\end{equation}
with $a_1 = 1, a_0 = 0, a_{\bar{1}} = -1$, where a hatted element in a tuple means that it is omitted.
Once we verify \eqref{Ti-equations}, the master equation \eqref{mastereq} holds because  $\sum_i T_i(\tau)$ telescopes to zero.

We now verify the equations in \eqref{Ti-equations} one at a time. We will start with the bulk equation for $T_i$ in \eqref{Ti-def} and use \eqref{bulkansatz} for all nine possibilities for $(\tau_i, \tau_{i+1})$. 
If $\tau_i = \tau_{i+1}$, there is nothing to check.
The other cases are verified in the following table.
\[
\begin{array}{c|c}
(\tau_i, \tau_{i+1}) & T_i(\tau) \\
\hline
(1,0) & \pi(\dots, 1,0, \dots) = \pi(\dots, 0, \dots) \\
(1, \bar{1}) & \pi(\dots, 1,\bar{1}, \dots) = \pi(\dots, \bar{1}, \dots) + \pi(\dots, 1, \dots) \\
(0, 1) & - \pi(\dots, 1,0, \dots) = - \pi(\dots, 0, \dots) \\
(0, \bar{1}) & \pi(\dots, 0,\bar{1}, \dots) = \pi(\dots, 0, \dots) \\
(\bar{1}, 1) & -\pi(\dots, 1,\bar{1}, \dots) = -\pi(\dots, 1, \dots) - \pi(\dots, \bar{1}, \dots) \\
(\bar{1}, 0) & -\pi(\dots, 0,\bar{1}, \dots) = -\pi(\dots, 0, \dots) \\
\hline
\end{array}
\]
For the $T_1$ equation, the nontrivial cases are analyzed using \eqref{leftansatz} and \eqref{bulkansatz}.
\[
\begin{array}{c|c}
(\tau_1, \tau_{2}) & T_1(\tau) \\
\hline
(*,1) & -\alpha \;  \pi(*,\bar{1}, \dots) = -\pi(*, \dots) \\
(*,0) & \alpha_*  \; \pi(*,0,\dots) - \pi(0,\bar{1},\dots) = 0 \\
(*,\bar{1}) & \alpha  \; \pi(*,\bar{1}, \dots) = \pi(*,\dots) \\
(0,1) & -\alpha_*  \; \pi(*,0, \dots) = - \pi(0,\dots) \\
(0,\bar{1}) & \pi(0,\bar{1},\dots) = \pi(0,\dots) \\
\hline
\end{array}
\]
Lastly, we consider the equation for $T_{n-1}$, which we analyze using \eqref{rightansatz} and \eqref{bulkansatz}. This can be justified using the charge-conjugation symmetry mentioned earlier, 
but it might be illustrative to write out all the details.
\[
\begin{array}{c|c}
(\tau_{n-1}, \tau_{n}) & T_{n-1}(\tau) \\
\hline
(1,*) & \beta  \; \pi(\dots, 1,*) = \pi(\dots, *)\\
(1,0) & \pi(\dots, 1,0) = \pi(\dots, 0)\\
(0,*) & \beta_*  \; \pi(\dots, 0,*) - \pi(\dots 1,0) = 0\\
(\bar{1},*)  & -\beta  \; \pi(\dots, 1,*) = -\pi(\dots, *)\\
(\bar{1},0)  & -\beta_*  \; \pi(\dots, 0,*) = - \pi(\dots, 0)\\
\hline
\end{array}
\]
In each case, we have shown that \eqref{Ti-equations} are satisfied, completing the proof of the matrix algebra.

To show that the algebra given by \eqref{bulkansatz}, \eqref{leftansatz} and
\eqref{rightansatz} is consistent, we give an explicit representation. 
We first choose the operators $X_+$ and $X_-$ as
\begin{equation}
\label{repn x+-}
X_+ = \begin{pmatrix}
1 & 1 & 0  & \dots \\
0 & 1 & 1  & \ddots \\
0 & 0 & 1 &  \ddots \\
\vdots & \ddots & \ddots &  \ddots
\end{pmatrix}, 
\quad
X_- = \begin{pmatrix}
1 & 0 & 0  & \dots \\
1 & 1 & 0  & \ddots \\
0 & 1 & 1 &  \ddots \\
\vdots & \ddots & \ddots &  \ddots
\end{pmatrix}.
\end{equation}
To build the representation, we recall the vectors $\bra{\widetilde{W}_\alpha}$
and $\ket{\widetilde{V}_\beta}$
\begin{equation}
\bra{\widetilde{W}_\alpha} = \left(1, \frac{1-\alpha}{\alpha}, \left( \frac{1-\alpha}{\alpha} \right)^2,\dots\right),
\quad
\ket{\widetilde{V}_\beta} = \begin{pmatrix}
1 \\
\ds \frac{1-\beta}{\beta} \\
\ds \left(\frac{1-\beta}{\beta} \right)^2\\
\vdots
\end{pmatrix},
\end{equation}
which satisfy $\bra{\widetilde{W}_\alpha} X_- = 1/\alpha \bra{\widetilde{W}_\alpha} $
and $X_+ \ket{\widetilde{V}_\beta} = 1/\beta \ket{\widetilde{V}_\beta}$. Using these vectors,
we write
\begin{equation}
\label{repn x0*}
X_0 = \begin{pmatrix}
1 & 0 & 0  & \dots \\
0 & 0 & 0  & \ddots \\
0 & 0 & 0 &  \ddots \\
\vdots & \ddots & \ddots &  \ddots
\end{pmatrix} = \ket{\widetilde{V}_1} \bra{\widetilde{W}_1}
\quad
\text{and}
\quad
X_* = \frac{1}{\alpha_* + \beta_*} \ket{\widetilde{V}_\beta} \bra{\widetilde{W}_\alpha}.
\end{equation}
Since $X_+, X_-$ and $X_0$ are written in the form of one of the standard representations, 
the bulk algebra \eqref{bulkansatz} is satisfied. 
Notice that $X_0$ satisfies
\begin{equation}
\label{x0 identities}
X_0^2 = X_0 = [X_+, X_-] = \ket{\widetilde{V}_1} \bra{\widetilde{W}_1}
\end{equation}
in this representation.
Finally, we set the vectors
\begin{equation}
\label{repn wv}
\begin{split}
\bra{W} &= \left(\alpha_*, \beta_* \frac{\beta}{2(1-\beta)}, \beta_* \left( \frac{\beta}{2(1-\beta)} \right)^2, \dots \right), \\
\ket{V} &= \begin{pmatrix}
\beta_* \\
\ds \alpha_* \frac{\alpha}{2(1-\alpha)} \\
\ds \alpha_* \left(\frac{\alpha}{2(1-\alpha)} \right)^2\\
\vdots
\end{pmatrix},
\end{split}
\end{equation}
so that $\bra{W} X_* = \bra{\widetilde{W}_\alpha}$ and $X_* \ket{V} = \ket{\widetilde{V}_\beta}$. 
One can use the above identities to verify that the 
boundary relations \eqref{leftansatz} and \eqref{rightansatz}
are satisfied.

\begin{example} 
Consider the \dstar{} with $n = 3$ and $n_0 = 1$. The configurations are
\[
\Omega_{3,1} = \{ (0, \bar{1}, *), (0, 1, *), (*, \bar{1}, 0), (*, 1, 0), (*, 0, *) \}.
\]
The column-stochastic generator of the \dstar{} in the ordered basis of $\Omega_{3,1}$ is then given by
\[
\left(
\begin{array}{ccccc}
 -1 & \beta & 0 & 0 & 0 \\
 0 & -\beta & 0 & \alpha_* & 0 \\
 0 & 0 & -\alpha & \beta_* & 0 \\
 1 & 0 & 0 & -\alpha_*-\beta_* & 1 \\
 0 & 0 & \alpha & 0 & -1 \\
\end{array}
\right).
\]
The stationary distribution is the column eigenvector of the matrix above with eigenvalue zero, and is given in the same basis by
\begin{equation}
\label{eg:ss31}
\frac{1}{Z_{3,1}} \left( \frac{1}{ \beta_*} , \frac{1}{\beta \beta_*},
\frac{1}{\alpha \alpha_* }, \frac{1}{ \alpha_* }, \frac{1}{\alpha_* \beta_*} \right),
\end{equation}
where the nonequilibrium partition function is 
\[
Z_{3,1} = \frac{1}{ \alpha_* } + 
\frac{1}{\alpha \alpha_* } +
\frac{1}{\alpha_* \beta_*} +
\frac{1}{\beta \beta_*} +
\frac{1}{ \beta_*} .
\]
One can easily check that the stationary probabilities computed by the matrix ansatz are given by \eqref{eg:ss31}.
\end{example}

\section{The Partition function}
\label{sec:partfn}

In this section, we will derive an explicit formula for the partition function $Z_{n,n_0}$. To do so, we will need to recall some notation. 
The $n$'th {\em Catalan number} is given by
\begin{equation}
\label{catalan-numbers}
\catalan{n} = \frac{1}{n+1} \binom {2n}n, \quad n \in \mathbb{N}.
\end{equation}
The Catalan numbers form an important sequence in enumerative combinatorics.
For example, the number of up-right paths
from $(0,0)$ to $(n,n)$ which stay on or below the diagonal $x=y$ are all counted by the
Catalan numbers.
See \cite[Sequence A000108]{oeis} for more details on the sequence.
Among the many refinements of the Catalan numbers, an important one is the sequence of 
{\em ballot numbers}, $\ballot{n}{k}$ given by
\begin{equation}
\label{ballot-numbers}
\ballot{n}{k} = \frac{n-k+1}{n+1} \binom {n+k}k, \quad 0 \leq k \leq n. 
\end{equation}
The ballot numbers $\ballot{n}{k}$ \cite[Sequence A009766]{oeis} count the number of up-right paths
from $(0,0)$ to $(n,n)$  which stay on or below the diagonal $x=y$ and which touch the diagonal exactly $n-k+1$ times (counting both endpoints). 

We will now show that the partition function is given by
\begin{equation}
\label{pf-dstar}
\begin{split}
Z_{n,n_0} = \ballot{n+n_0-3}{n-n_0} + \sum_{k=0}^{n-n_0-1} \ballot{n+n_0-3}k & \left(  
\frac{1}{\alpha^{n-n_0-1-k} \alpha_*} + \frac{1}{\beta_* \beta^{n-n_0-1-k}} 
\right. \\
&+ \sum_{j=0}^{n-n_0-2-k} 
\left. \frac{1}{\alpha_* \beta_* \alpha^{j} \beta^{n-n_0-2-k-j}} \right).
\end{split}
\end{equation}
As a step towards the proof of \eqref{pf-dstar}, we consider the set
of configurations for the semipermeable TASEP, 
\[
\hat{\Omega}_{n,n_0} = \{ \tau \in \{\1,0,1\}^n \; | \; n_0(\tau) = n_0 \},
\]
and let
\begin{equation}
\label{defF}
\hat{F}_{n,n_0} = \sum_{\tau \in \hat{\Omega}_{n,n_0}} X_{\tau_1} \dots
X_{\tau_n}.
\end{equation}
It has been shown by Arita (see \cite[Appendix A]{arita-2006}) that
\begin{equation}
\label{pf-typeC}
\hat{F}_{n,n_0} = \sum_{k=0}^{n-n_0}  \ballot{n+n_0-1}{n-n_0-k} \sum_{j=0}^k X_-^j X_0^{n_0}
X_+^{k-j}.
\end{equation}
Let us consider the sum we want to compute, 
\begin{equation}
\label{defF*}
F_{n,n_0} = \sum_{\tau \in \Omega_{n,n_0}} X_{\tau_1} \dots
X_{\tau_n}.
\end{equation}
Using \eqref{defF}, we see that
\begin{align}
\label{F*expand}
F_{n,n_0} = X_* \hat{F}_{n-2,n_0} X_* + X_* \hat{F}_{n-2,n_0-1} X_0 + X_0 \hat{F}_{n-2,n_0-1} X_* + X_0 \hat{F}_{n-2,n_0-2} X_0.
\end{align}
Substitute $\hat{F}_{n,n_0}$ from \eqref{pf-typeC} in the above expression. The first term in \eqref{F*expand} becomes
\begin{equation}
\label{pf-1st term}
\sum_{k=0}^{n-n_0-2}  \ballot{n+n_0-3}{n-n_0-2-k} \sum_{j=0}^k 
X_* X_-^j X_0^{n_0} X_+^{k-j} X_*.
\end{equation}
The second term in \eqref{F*expand} simplifies to
\begin{equation}
\label{pf-2nd term}
\begin{split}
\sum_{k=0}^{n-n_0-1}  \ballot{n+n_0-4}{n-n_0-1-k} \sum_{j=0}^k 
X_* X_-^j X_0^{n_0-1} X_+^{k-j} X_0
= & \sum_{k=0}^{n-n_0-1}  \ballot{n+n_0-4}{n-n_0-1-k} \sum_{j=0}^k 
X_* X_-^j X_0^{n_0} \\
= & \sum_{j=0}^{n-n_0-1}  X_* X_-^j X_0^{n_0} 
\sum_{k=j}^{n-n_0-1} \ballot{n+n_0-4}{n-n_0-1-k} \\
=& \sum_{j=0}^{n-n_0-1} \ballot{n+n_0-3}{n-n_0-1-j}  X_* X_-^j X_0^{n_0}, 
\end{split}
\end{equation}
where we have used the standard identity,
\begin{equation}
\label{cattri-id1}
\sum_{a=0}^b \ballot ma = \ballot{m+1}b,
\end{equation} 
in the last equality.
The third term in \eqref{F*expand} simplifies in a similar manner to 
\begin{equation}
\label{pf-3rd term}
\sum_{j=0}^{n-n_0-1} \ballot{n+n_0-3}{n-n_0-1-j}  X_0^{n_0} X_+^j X_*.
\end{equation}
Finally, the fourth term in \eqref{F*expand} yields
\begin{equation}
\label{pf-4th term}
\begin{split}
\sum_{k=0}^{n-n_0}  \ballot{n+n_0-5}{n-n_0-k} \sum_{j=0}^k 
X_0 X_-^j X_0^{n_0-2} X_+^{k-j} X_0 
= \sum_{k=0}^{n-n_0}  (k+1) \ballot{n+n_0-5}{n-n_0-k} X_0^{n_0} 
= \ballot{n+n_0-3}{n-n_0} X_0^{n_0},
\end{split}
\end{equation}
where we have used the identity
\begin{equation}
\label{cattri-id2}
\sum_{k=0}^b (k+1) \ballot m{b-k} = \ballot{m+2}b,
\end{equation}
in the last equality.
To summarise, we have now shown
\begin{align*}
F_{n,n_0} = & \sum_{k=0}^{n-n_0-2}  \ballot{n+n_0-3}{n-n_0-2-k} \sum_{j=0}^k X_* X_-^j X_0^{n_0} X_+^{k-j} X_* \\
& + \sum_{j=0}^{n-n_0-1} \ballot{n+n_0-3}{n-n_0-1-j} ( X_* X_-^j X_0^{n_0} + X_0^{n_0} X_+^j X_*) + \ballot{n+n_0-3}{n-n_0} X_0^{n_0}.
\end{align*}
Using the boundary algebra \eqref{leftansatz} and \eqref{rightansatz} and properties of the representation \eqref{repn x+-}, \eqref{repn x0*} and \eqref{repn wv} completes the proof of \eqref{pf-dstar}.

In the special case when $\alpha = \alpha_* = \beta = \beta_* = 1$, it can be shown that the partition function in \eqref{pf-dstar} reduces to
$Z_{n,n_0} = \ballot{n+n_0-1}{n-n_0}$ when $n_0 > 0$, and $Z_{n,0} = \catalan{n-1}$.

\section{Current and Density}
\label{sec:curr-dens}

As the number of $0$'s is conserved, their current is zero. The current of $1$'s to the right is the same as that of $\bar{1}$'s to the left. Since all species of particles are conserved in the bulk, the current is independent of the site. We denote the current of $1$'s by $J$.

By the matrix ansatz description of stationary probabilities in \eqref{statprob}, the rate at which a $1$ hops across the $(i,i+1)$'th bond is given by
\begin{equation}
\begin{split}
\label{current-matrixansatz}
J =& \frac{1}{Z_{n,n_0}} \left( \sum'_{\tau \in \Omega_{n,n_0}} \bra W X_{\tau_1} \dots X_{\tau_{i-1}} X_+ X_0 X_{\tau_{i+2}} \dots X_{\tau_n}  \ket V \right. \\
&+ \left. \sum''_{\tau \in \Omega_{n,n_0}} \bra W X_{\tau_1} \dots X_{\tau_{i-1}} X_+ X_- X_{\tau_{i+2}} \dots X_{\tau_n}  \ket V \right),
\end{split}
\end{equation}
where the sums $\sum'$ (resp. $\sum''$) runs over all configurations $\tau$ with $\tau_i = 1$, $\tau_{i+1} = 0$ (resp. $\tau_{i+1} = \bar{1}$), and  $n_0$ (resp. $n_0-1$) $0$'s among sites in $\{1,\dots,i-1,i+2,\dots,n\}$. Using the matrix algebra in \eqref{bulkansatz}, we simplify \eqref{current-matrixansatz} to
\[
\begin{split}
J =& \frac{1}{Z_{n,n_0}} \left( \sum'_{\tau \in \Omega_{n,n_0}} \bra W X_{\tau_1} \dots X_{\tau_{i-1}} X_0 X_{\tau_{i+2}} \dots X_{\tau_n}  \ket V \right. \\
&+ \left. \sum''_{\tau \in \Omega_{n,n_0}} \bra W X_{\tau_1} \dots X_{\tau_{i-1}} (X_+ + X_-) X_{\tau_{i+2}} \dots X_{\tau_n}  \ket V \right),
\end{split}
\]
and it is easy to see that this simplifies to
\begin{equation}
\label{current-formula}
J = \frac{Z_{n-1,n_0}}{Z_{n,n_0}}.
\end{equation}

We now give a formula for the density. It will suffice to give a formula for the density of $1$'s for the following reason: By the charge-conjugation symmetry mentioned in Section~\ref{sec:intro}, we will obtain a formula for the density of $\bar{1}$'s and since the total density at any site is $1$, we will obtain the density of $0$'s.
Let $\rho^+_i$ denote the density of $1$'s at the $i$'th site in the steady state, i.e., the probability that there is a $1$ at site $i$ in the stationary distribution. Similarly, let $\rho^-_i$ and $\rho^0_i$ denote the densities of $\bar 1$'s and $0$'s at site $i$ respectively.
Let $G_\mathsf{y} = X_+ + \mathsf{y} X_0 + X_-$, where $\mathsf{y}$ is a formal parameter. 
By the matrix ansatz \eqref{statprob},
\begin{equation}
\label{dens-matrix ansatz}
\rho^+_i = \frac{1}{Z_{n,n_0}} [\mathsf{y}^{n_0}]
\bra W (X_* + \mathsf{y} X_0) G_\mathsf{y}^{i-2} X_+ G_\mathsf{y}^{n-i-1} (X_* + \mathsf{y} X_0)
\ket V,
\end{equation}
where the expression $[x^a] P(x)$ denotes the coefficient of $x^a$ in $P(x)$. 

We first give a formula for $\rho^+_i$ when $2 \leq i \leq n-3$. 
Arita has shown \cite[Eq. (33)]{arita-2006} that, for $j \geq 1$,
\[
X_+ G_\mathsf{y}^j = \sum_{k=0}^{j-1} \catalan{k} G_\mathsf{y}^{j-k}
+ \sum_{k=1}^j \ballot{j-1}{j-k} X_+^{k+1},
\]
where we recall the Catalan numbers and ballot numbers defined in \eqref{catalan-numbers} and \eqref{ballot-numbers} respectively. Using the above equation, \eqref{dens-matrix ansatz} splits into two terms. The first one 
is similar to the first sum in the  density formula for the semipermeable TASEP~\cite[Eq. (38)]{arita-2006}
and becomes
\[
\frac{1}{Z_{n,n_0}} [\mathsf{y}^{n_0}] \sum_{k=0}^{n-i-2} \catalan{k}
\bra W (X_* + \mathsf{y} X_0) G_\mathsf{y}^{n-k-3} (X_* + \mathsf{y} X_0)
\ket V,
\]
which immediately simplifies using \eqref{pf-dstar-matrix ansatz} to
\[
\sum_{k=0}^{n-i-2} \catalan{k} \frac{Z_{n-k-1,n_0}}{Z_{n,n_0}}.
\]
The second term in \eqref{dens-matrix ansatz} has to be treated more carefully. Using \eqref{rightansatz} for $X_+$, it becomes
\begin{equation}
\label{dens-second}
\begin{split}
\frac{1}{Z_{n,n_0}} [\mathsf{y}^{n_0}] \sum_{k=1}^{n-i-1} & \ballot{n-i-2}{n-i-1-k}  \Bigg(  \frac{1}{\beta^{k+1}} 
\bra W (X_* + \mathsf{y} X_0) G_\mathsf{y}^{i-2} X_* \ket V \\
&+  \mathsf{y} \bra W (X_* + \mathsf{y} X_0) G_\mathsf{y}^{i-2} X_0 \ket V \Bigg).
\end{split}
\end{equation}
To evaluate this, it will be convenient to define 
\begin{align*}
F^*_{n,n_0} &= X_* \hat{F}_{n-2,n_0} X_* + X_0 \hat{F}_{n-2,n_0-1} X_* , &\quad 
Z^*_{n,n_0} =  \bra W F^*_{n,n_0} \ket V, \\
F^0_{n,n_0} &= X_* \hat{F}_{n-2,n_0-1} X_0 + X_0 \hat{F}_{n-2,n_0-2} X_0, & \quad 
Z^0_{n,n_0} =  \bra W F^0_{n,n_0} \ket V,
\end{align*}
where the superscripts denote the particle on the last site.
It is then clear that $F_{n,n_0} = F^*_{n,n_0} + F^0_{n,n_0}$. 
We then obtain $F^*_{n,n_0}$ by adding \eqref{pf-1st term} and \eqref{pf-3rd term}, and $F^0_{n,n_0}$ by adding \eqref{pf-2nd term} and \eqref{pf-4th term}. 
Using these expressions, \eqref{dens-second} becomes
\begin{align*}
\frac{1}{Z_{n,n_0}} \sum_{k=1}^{n-i-1} \ballot{n-i-2}{n-i-1-k}
&  \bra W \Bigg( \frac{1}{\beta^{k+1}} F^*_{i,n_0} +   F^0_{i,n_0}
\Bigg) \ket V .
\end{align*}
Now, from the definitions,
\begin{equation}
\label{pf-sum-1}
Z^*_{n,n_0} = \sum_{k=0}^{n-n_0-1} \left( \ballot{n+n_0-3}{n-n_0-2-k} \sum_{j=0}^{n-n_0-2-k}  \frac{1}{\alpha_* \beta_* \alpha^{j} \beta^{k-j}}
+  \ballot{n+n_0-3}{n-n_0-1-k} \frac{1}{\beta_* \beta^{k}} \right),
\end{equation} 
and
\begin{equation}
\label{pf-sum-2}
Z^0_{n,n_0} = \sum_{k=0}^{n-n_0-1} \ballot{n+n_0-3}{n-n_0-1-k}  
\frac{1}{\alpha_* \alpha^{k}} + \ballot{n+n_0-3}{n-n_0},
\end{equation}
using which \eqref{dens-second} can be evaluated immediately. 
To summarise, we have shown that the density at site $i$, for $2 \leq i \leq n-2$, is given by
\begin{equation}
\label{dens-formula-generic}
\begin{split}
\rho^+_i = \sum_{k=0}^{n-i-2} \catalan{k} \frac{Z_{n-k-1,n_0}}{Z_{n,n_0}}
+ \frac{Z^*_{i,n_0}}{Z_{n,n_0}} \sum_{k=1}^{n-i-1} \ballot{n-i-2}{n-i-1-k}
\frac{1}{\beta^{k+1}}  + \catalan{n-i-1} \frac{Z^0_{i,n_0}}{Z_{n,n_0}}.
\end{split}
\end{equation}
Notice that when $i < n_0$, the first sum in \eqref{dens-formula-generic} becomes independent of $i$ and the last two sums drop out. Therefore, the densities at all sites between $2$ and $n_0-1$ are identical. A similar result holds for the semipermeable TASEP~\cite{als-2009}. See Section~\ref{sec:discuss} for a discussion of this phenomenon.

There's a slightly different formula for the density at the penultimate site. Plugging in $i=n-1$ in \eqref{dens-matrix ansatz}, we obtain
\begin{align*}
\rho^+_{n-1} =& \frac{1}{Z_{n,n_0}} [\mathsf{y}^{n_0}]
\bra W (X_* + \mathsf{y} X_0) G_\mathsf{y}^{n-3} X_+ (X_* + \mathsf{y} X_0)
\ket V \\
=& \frac{1}{Z_{n,n_0}} [\mathsf{y}^{n_0}]
\bra W (X_* + \mathsf{y} X_0) G_\mathsf{y}^{n-3} (\frac{1}{\beta} X_* + \mathsf{y} X_0)
\ket V,
\end{align*}
and simplifying using \eqref{pf-sum-1} and \eqref{pf-sum-2}, we find
\begin{equation}
\label{dens-formula-last}
\rho^+_{n-1} = \frac{1}{Z_{n,n_0}} \left(
\frac{1}{\beta} Z^*_{n-1,n_0} + Z^0_{n-1,n_0} \right).
\end{equation}

\section{Phase diagram}
\label{sec:phase}

The steady state phase diagram of the \dstar{} in the thermodynamic limit depends on three parameters, $\alpha, \beta$ and the density of $0$'s. We take $n, n_0 \to \infty$ so that there is a limiting density of $0$'s, i.e. $n_0/n \to \zeta$, where $0 \leq \zeta \leq 1$. 
Therefore, our phase diagram is three-dimensional and will depend on three parameters, $\alpha, \beta$ and $\zeta$, each of which lies between $0$ and $1$. It will be convenient for us to consider the $(\alpha, \beta)$-plane at a fixed value of $\zeta$.
Note that unlike the rest of the paper, some computations in this section are mathematically non-rigorous.

\begin{figure}[htbp!]
		\begin{center}	
			\begin{tikzpicture}[scale=0.5]
			\draw[black,thick] (0,0)--(3,3);
			\draw[black,thick] (0,0) rectangle (10,10);
			\draw[black,thick] (3,10)--(3,3);
			\draw[black,thick] (10,3)--(3,3);
			
			\node at (6.5,6.5) {I};
			\node at (1.5,6.5) {II};
			\node at (6.5,1.5) {III};
			
			\node at (-0.65,0) {$0$};
			\node at (-0.65,10) {$1$};
			\draw (0,3) -- (0.2,3);
			\node at (-0.75,3) {$\frac{1-\zeta}{2}$};
			\node at (-1,5) {$\beta$};
			
			\node at (0,-0.65) {$0$};
			\node at (10,-0.65) {$1$};
			\draw (3,0) -- (3,0.2);
			\node at (3,-0.75) {$\frac{1-\zeta}{2}$};
			\node at (5,-1) {$\alpha$};

			\end{tikzpicture}
		\end{center}
	\caption{The phase diagram for the \dstar{}.}
	\label{fig:phase-diag}
\end{figure}
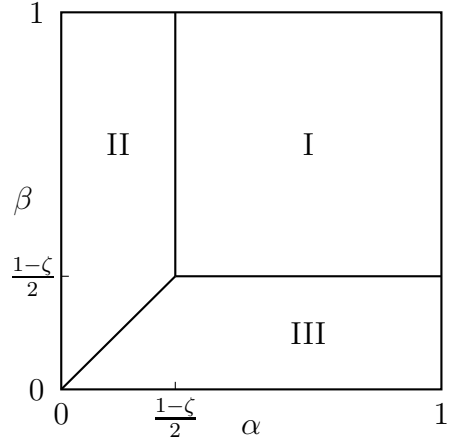

We will now determine the phase diagram using the current as the order parameter.
Recall that the current is given by \eqref{current-formula} in the steady state. Therefore, we need leading order asymptotics for the partition function $Z_{n,\zeta n}$. These have been computed in Appendix~\ref{sec:asymp} following a similar strategy as in \cite{arita-2006}. 
Using the results in \eqref{pf-asymp-1}--\eqref{pf-asymp-5}, we find that
\[
J = \frac{1- \zeta^2}{4} \quad \text{when } \alpha, \beta \geq 
\frac{1-\zeta}{2}.
\]
We will call this region of the phase diagram {\em Phase I}.
Similarly, \eqref{pf-asymp-6} shows that
\[
J = \alpha(1-\alpha) \quad \text{when } \alpha < \beta, \alpha < (1-\zeta)/2.
\]
This region is denoted {\em Phase II}.
Lastly, \eqref{pf-asymp-7}--\eqref{pf-asymp-8} show that
\[
J = \beta(1-\beta) \quad \text{when } \beta \leq \alpha, \beta < (1-\zeta)/2.
\]
This region is denoted {\em Phase III}.
The phase diagram is identical to that of the semipermeable TASEP~\cite{arita-2006,als-2009} and is shown in Figure~\ref{fig:phase-diag}.
All transitions are of second order, since the current is continuous, but not differentiable across all phase boundaries.

We now compute the density of $1$'s far from the boundaries in the large volume limit. We fix a position $i = xn$ for $x \in (0,1)$ and $n_0 = \zeta n$, and take the limit as $n \to \infty$ in \eqref{dens-formula-generic}. 
For the first sum, write $Z_{n,\zeta n} = C \lambda^n \, \mu^{\zeta n} \, n^z$, where $C, \lambda, \mu, z$ depend on the phases as shown in \eqref{pf-asymp-1}--\eqref{pf-asymp-8}. Then we get
\[
\sum_{k=0}^{n(1-x)-2} C_k \frac1{\lambda^{k+1}}
\approx \frac{1 - \sqrt{1-4 /\lambda}}{2}
= \begin{cases}
\ds \frac{1-\zeta}{2} & \text{Phase I}, \\[0.2cm]
\alpha & \text{Phase II}, \\
\beta & \text{Phase III}. \\
\end{cases}
\]
This is a place where the computation is mathematically non-rigorous and we leave it as a challenge for the interested reader to make it rigorous.

We now consider the second term in \eqref{dens-formula-generic}.
Note that it is nonzero only when $x \geq \zeta$. Again, write $Z^*_{n,\zeta n} = C_1 \lambda_1^n \, \mu_1^{\zeta n} \, n^{z_1}$ where $C_1, \lambda_1, \mu_1, z$ depend on the phases as shown in \eqref{pf1-asymp-1}--\eqref{pf1-asymp-8}. 
Moreover, we have
\[
\sum_{k=1}^{n-i-1} \ballot{n-i-2}{n-i-1-k} \frac{1}{\beta^{k+1}}
= R_{n(1-x)-1,0}(\beta).
\]
The asymptotics of $R_{n,0}(\beta)$ have been computed in \cite[Eqs. (48)--(50)]{dehp} to be
\begin{equation}
\label{Rn0-asymp}
R_{n,0}(\beta) \approx
\begin{cases}
\ds \frac{1}{(2\beta-1)^2 \sqrt{\pi}} \frac{4^n}{n^{3/2}} 
& \beta > \frac{1}{2}, \\[0.4cm]
\ds \frac{2}{\sqrt{\pi}} \frac{4^n}{n^{1/2}}
& \beta = \frac{1}{2}, \\[0.4cm]
\ds (1-2\beta) \left( \beta(1-\beta) \right)^{-n-1}
& \beta < \frac{1}{2}.
\end{cases}
\end{equation}
Therefore, the asymptotics of the second term are given by
\[
\frac{C_1}{C} \frac{\lambda_1^{x n}}{\lambda^n} \frac{(x n)^z}{n^z}
\frac{\mu_1^{\zeta n}}{\mu^{\zeta n}} R_{n-i-1,0}(\beta)
= \frac{C_1}{C} \frac{\lambda_1^{x n}}{\lambda^n} x^z R_{n(1-x)-1,0}(\beta).
\]
The only regions in which this term does not go to zero is when the exponential terms in $R$ match those of the prefactor. This can only happen if both $Z_{n,\zeta n}$ and $Z^*_{n,\zeta n}$ lie in either phase III or the II-III boundary. Therefore, we must have $\beta \leq \alpha$ and $\beta \leq \frac{1-\zeta/x}{2}$. 

First, suppose $\beta < \alpha$.
In case $\beta < \frac{1-\zeta/x}{2}$, using \eqref{pf-asymp-7} and \eqref{pf1-asymp-7}, we find that $C_1 = C$, $z=z_1 = 0$, $\lambda_1 = \lambda = 1/(\beta(1-\beta))$
and $\mu_1 = \mu = \beta/(1-\beta)$. Further, using \eqref{Rn0-asymp}, we find that the second term becomes $1-2\beta$.
On the other hand, if $\beta = \frac{1-\zeta/x}{2}$, $C_1 = C/2$ and the rest is as before, so that the second term is $(1-2\beta)/2$.

Now, if $\beta = \alpha$, we must have $\beta < \frac{1-\zeta/x}{2}$ for a nontrivial contribution. In that case, using \eqref{pf-asymp-8} and \eqref{pf1-asymp-8}, we find that $z=z_1 = 1$, $\lambda_1 = \lambda = 1/(\beta(1-\beta))$, $\mu_1 = \mu = \beta/(1-\beta)$ and
\[
\frac{C_1}{C} = \frac{1 - \zeta/x - 2 \beta}{1 - \zeta - 2\beta}.
\]
Taking into account the asymptotics from \eqref{Rn0-asymp}, we find that the second term becomes
\[
(1-2\beta) \frac{x(1 - \zeta/x - 2 \beta)}{1 - \zeta - 2\beta}
= \frac{(1-2\beta)^2}{1 - \zeta - 2\beta}
\left(x - \frac{\zeta}{1-2\beta} \right).
\]

Lastly, we consider the third term in \eqref{dens-formula-generic}. 
Write $Z^0_{n,\zeta n} = C_2 \lambda_2^n \mu_2^{\zeta n} n^{z_2}$ where $C_2, \lambda_2, \mu_2, \allowbreak z$ depend on the phases as shown in \eqref{pf2-asymp-1}--\eqref{pf2-asymp-3}. Similar to the second term, we need to consider the ratio of $Z^0_{x n,\zeta n}$ and $Z^*_{n,\zeta n}$ in various phases. It is clearly nonzero only if $x \geq \zeta$. The prefactor of the Catalan number has asymptotics given by
\[
\catalan{n-i-1} \approx \frac{4^{n(1-x)-1}}{\sqrt{\pi}\, (n(1-x))^{3/2}}.
\]
A little thought shows that in all regions of the phase diagram the third term goes to zero exponentially in $n$, and thus does not contribute to the density in the thermodynamic limit. Therefore, the density is given by the contribution of the first and second terms in \eqref{dens-formula-generic}. 

We now summarise the asymptotics of the density formulas using the charge conjugation symmetry explained in Section~\ref{sec:intro}. Let $x_1 =  \frac{\zeta}{1-2\alpha}$ and $x_2 = \frac{\zeta}{1-2\beta}$. Let $\rho^+(x), \rho^0(x), \allowbreak \rho^-(x)$ denote the densities of $1$'s, $0$'s and $\bar{1}$'s respectively at position $\lfloor x n \rfloor$ for $x \in [0,1]$ as $n \to \infty$.
The densities in various regions are given in Table~\ref{tab:dens-asymp}. 

\begin{table}[htb!]
\begin{tabular}{|l|c|c|c|}
\hline
&&&\\[-0.3cm]
Phase & $\rho^+(x)$ & $\rho^0(x)$ & $\rho^-(x)$ \\
\hline 
&&&\\[-0.1cm]
I & $\ds \frac{1-\zeta}{2}$ & $\zeta$ & $\ds \frac{1-\zeta}{2}$ \\[0.3cm]
\hline
&&&\\[-0.1cm]
\multirow{3}{*}{II} & \multirow{3}{*}{$\alpha$} 
& $0$ for $x < 1-x_1$, &  $1-\alpha$ for $x < 1-x_1$, \\[0.3cm]
&& $\ds \frac{1}2$ for $x = 1-x_1$, 
& $\ds \frac{1-2\alpha}2$ for $x = 1-x_1$, \\[0.3cm]
&& $1-2\alpha$ for $x > 1-x_1$. & $\alpha$ for $x > 1-x_1$. \\
\hline
&&&\\[-0.1cm]
\multirow{3}{*}{III} & $\beta$ for $x < x_2$, & 
$1-2\beta$ for $x < x_2$, &\multirow{3}{*}{$\beta$} \\[0.3cm]
& $\ds\frac{1}2$ for $x = x_2$, 
& $\ds\frac{1-2\beta}2$ for $x = x_2$, & \\[0.3cm]
& $1-\beta$ for $x > x_2$. 
& $0$ for $x > x_2$. & \\
\hline
&&&\\[-0.1cm]
II-III & $\alpha$ for $x \leq x_1$, 
& linear for $x < x_1, x> 1-x_1$, & linear for $x < 1-x_1$,\\[0.3cm]
boundary & linear for $x > x_1$. & $1-2\alpha$ for $x_1 \leq x \leq 1-x_1$.
& $\alpha$ for $x \geq 1-x_1$. \\
\hline
\end{tabular}
\vspace{0.3cm}
\caption{Densities of all three species in the bulk in  different phases in the thermodynamic limit.
The precise formulas for the linear profiles on the II-III boundary are the same as in~\cite{arita-2006} and can be seen from the discussion in Section~\ref{sec:phase}.}
\label{tab:dens-asymp}
\end{table}

\section{Discussion}
\label{sec:discuss}

The motivation for this work comes from Lam's study on random walks for affine Weyl groups~\cite{lam-2015}. The random walk for the affine Weyl group of type A is directly related to the multispecies TASEP on the ring. The matrix ansatz for the latter was determined in \cite{evans-ferrari-mallick-2009} and the steady state was determined exactly using multiline queues in \cite{FM07}. In both these studies, the understanding of the two-species case (\cite{djls} for the matrix ansatz and \cite{angel-2006} for the queueing picture) was essential in building the general theory. In this work, we have considered the two-species TASEP whose study is crucial for the understanding of the random walk on the affine Weyl group of type D. In future work~\cite{AALP-2019},  
we plan to study the multispecies version of this TASEP. 
We also remark that we have chosen the rates so that we can study the random walks for types B and C as special cases.

The physics of the \dstar{} is very similar to that of the semipermeable TASEP discussed in \cite{als-2009}. In particular, the asymptotics strongly suggests that the phase diagram can be explained in terms of the {\em fat shock} defined in \cite{als-2009}. Recall that the fat shock is a bound state of all the $0$'s in the system. This fat shock gets pinned to the right boundary in Phase II, to the left boundary in Phase III, and pervades the system in Phase I. On the II-III boundary, the two fronts of the fat shock perform a correlated unbiased random walk, leading to piecewise linear density profiles for all species.

We also expect other phenomena for the semipermeable TASEP to continue to hold. For example, in the finite semipermeable TASEP, the joint correlation of $1$'s at sites $i_1 <\cdots < i_k <n_0$ only depended on $k$ and not the positions themselves. We have only proven a similar result for the \dstar{} for $k=1$ here, but exact computations for small sizes suggests that a similar result should be true for all $k < n_0-1$ here. If so, then the property of exchangeability of the limiting measure as seen from the left boundary should also continue to hold.

\section*{Acknowledgements}
AA was partially supported by the UGC Centre for Advanced Studies and by Department of Science and Technology grants DST/INT/SWD/VR/P-01/2014 and EMR/2016/ 006624.
SL and SP were supported by the Swedish Research Council grant 621-2014-4780.

\newcommand{\etalchar}[1]{$^{#1}$}
\providecommand{\bysame}{\leavevmode\hbox to3em{\hrulefill}\thinspace}
\providecommand{\MR}{\relax\ifhmode\unskip\space\fi MR }
\providecommand{\MRhref}[2]{%
  \href{http://www.ams.org/mathscinet-getitem?mr=#1}{#2}
}
\providecommand{\href}[2]{#2}

\appendix

\section{Asymptotic computations}
\label{sec:asymp}

To compute the asymptotics, it is useful to rewrite the partition function $Z_{n,n_0}$ in \eqref{pf-dstar} in a more compact way.  To that end, define
\begin{equation}
\label{defR}
R_{n,n_0}(\alpha) = \sum_{k=0}^{n-n_0} \ballot{n+n_0-1}{n-n_0-k} \frac{1}{\alpha^{k+1}}
\end{equation}
and
\begin{equation}
\label{defS}
S_{n,n_0}(\alpha) = \sum_{k=0}^{n-n_0} (k+1) \ballot{n+n_0-1}{n-n_0-k} \frac{1}{\alpha^{k}}.
\end{equation}
Then the second term on the right hand side of \eqref{pf-dstar} can be written as
\begin{align*}
\sum_{k=0}^{n-n_0-1} \ballot{n+n_0-3}k \frac{1}{\alpha^{n-n_0-1-k} \alpha_*} 
&= \ballot{n+n_0-3}{n-n_0-1} \frac{1}{\alpha_*} + 
\sum_{k=1}^{n-n_0-1} \ballot{n+n_0-3}{n-n_0-1-k} \frac{1}{\alpha^k \alpha_*} \\ 
&= \frac{1}{\alpha_*} \left( \ballot{n+n_0-3}{n-n_0-1}  +  
R_{n-2,n_0}(\alpha) \right),
\end{align*}
and similarly for the third term. Finally, if $\alpha \neq \beta$, the last term of \eqref{pf-dstar}
is
\begin{align*}
\sum_{k=0}^{n-n_0-2} \sum_{j=0}^{n-n_0-2-k} \ballot{n+n_0-3}{k}
 \frac{1}{\alpha_* \beta_* \alpha^{j} \beta^{n-n_0-2-k-j}}  
  &= \frac{1}{\alpha_* \beta_*} \sum_{k=0}^{n-n_0-2} 
 \ballot{n+n_0-3}{n-n_0-2-k} \sum_{j=0}^{k}
 \frac{1}{\alpha^{j} \beta^{k-j}} \\
 &= \frac{\alpha \beta}{\alpha_* \beta_* (\beta - \alpha)}
  \left( R_{n-2,n_0} (\alpha) - R_{n-2,n_0} (\beta) \right).
\end{align*}
If $\alpha = \beta$, using the previous computation, the last term of \eqref{pf-dstar}
is
\begin{align*}
\frac{1}{\alpha_* \beta_*} \sum_{k=0}^{n-n_0-2} 
 \ballot{n+n_0-3}{n-n_0-2-k} \beta^{-k} \sum_{j=0}^{k} 
 \left( \frac{\beta}{\alpha} \right)^j 
 &= \frac{1}{\alpha_* \beta_*} \sum_{k=0}^{n-n_0-2} 
 \ballot{n+n_0-3}{n-n_0-2-k} \beta^{-k} \sum_{j=0}^{k} 
 1\\
 &= \frac{1}{\alpha_* \beta_*} \sum_{k=0}^{n-n_0-2} (k+1)
 \ballot{n+n_0-3}{n-n_0-2-k} \beta^{-k} \\
 &= \frac{1}{\alpha_* \beta_*}S_{n-2, n_0}\left( \alpha \right).
\end{align*}
To summarise, we have obtained
\begin{equation}
\label{Z*-R}
\begin{split}
Z_{n,n_0} =\ &\ballot{n+n_0-3}{n-n_0} + \ballot{n+n_0-3}{n-n_0-1}
\left( \frac{1}{\alpha_*} + \frac{1}{\beta_*} \right) 
+ \frac{1}{\alpha_*} R_{n-2,n_0}(\alpha) 
+ \frac{1}{\beta_*} R_{n-2,n_0}(\beta) \\
&+ \frac{\alpha \beta}{\alpha_* \beta_* (\beta - \alpha)}
  \left( R_{n-2,n_0} (\alpha) - R_{n-2,n_0} (\beta) \right),
\end{split}
\end{equation} 
if $\alpha \neq \beta$, and
\begin{equation}
\label{Z*-RS}
Z_{n,n_0} =\ \ballot{n+n_0-3}{n-n_0} + \left(\ballot{n+n_0-3}{n-n_0-1} + R_{n-2,n_0}(\alpha)\right)
\left( \frac{1}{\alpha_*} + \frac{1}{\beta_*} \right) 
+ \frac{1}{\alpha_* \beta_*}S_{n-2, n_0}\left( \alpha \right),
\end{equation} 
if $\alpha = \beta$.

The advantage of writing the partition function this way is that the leading order asymptotics for $R_{n,n_0}(\alpha)$ and $S_{n,n_0}(\alpha)$ has been computed by Arita. We now describe his computation. Let $n, n_0 \rightarrow \infty$ so that $n_0/n \rightarrow \zeta$, where $0 < \zeta < 1$ is the (fixed) density of zeros in the system.
We will use the notation $a_n \approx b_n$ to mean that the ratio $a_n/b_n \to 1$ as $n \to \infty$.
 Then we have~\cite[Appendix B.1]{arita-2006} 
\begin{equation}
\label{R-asymp}
\begin{split}
& \ds R_{n, \zeta n} (\alpha) \approx 
\begin{cases}
\ds \frac{2 \zeta}{(2 \alpha - 1 + \zeta) \sqrt{\pi (1- \zeta^2) n}}
\left( \frac{4}{1-\zeta^2} \right)^n \left( \frac{1-\zeta}{1+\zeta} \right)^{\zeta n} &
\ds \alpha > \frac{1-\zeta}{2}, \\[0.5cm]
\ds \frac{2\zeta}{1-\zeta^2} 
\left( \frac{4}{1-\zeta^2} \right)^n \left( \frac{1-\zeta}{1+\zeta} \right)^{\zeta n} &
\ds \alpha = \frac{1-\zeta}{2}, \\[0.5cm]
\ds \frac{1 - 2\alpha}{\alpha(1-\alpha)} 
\left(\frac{1}{\alpha (1-\alpha)} \right)^n \left( \frac{\alpha}{1-\alpha} \right)^{\zeta n} &
\ds \alpha < \frac{1-\zeta}{2},
\end{cases}
\end{split}
\end{equation}
Note that there is a typo in \cite{arita-2006} for the case when $\alpha = (1-\zeta)/2$.
Similarly, we have ~\cite[Appendix B.2]{arita-2006}
\begin{equation}
\label{S-asymp}
\begin{split}
& \ds S_{n, \zeta n} (\alpha) \approx 
\begin{cases}
\ds \frac{4 \zeta\alpha^2}{(2 \alpha - 1 + \zeta)^2 \sqrt{\pi (1- \zeta^2) n}}
\left( \frac{4}{1-\zeta^2} \right)^n \left( \frac{1-\zeta}{1+\zeta} \right)^{\zeta n} & 
\ds \alpha > \frac{1-\zeta}{2}, \\[0.5cm]
\ds 2\zeta\sqrt{\frac{(1-\zeta)n}{\pi(1+\zeta)^3}}
\left( \frac{4}{1-\zeta^2} \right)^n \left( \frac{1-\zeta}{1+\zeta} \right)^{\zeta n} &
\ds \alpha = \frac{1-\zeta}{2}, \\[0.5cm]
\ds \frac{(1 - 2\alpha)(1-\zeta-2\alpha)n}{(1-\alpha)^2} 
\left(\frac{1}{\alpha (1-\alpha)} \right)^n 
\left( \frac{\alpha}{1-\alpha} \right)^{\zeta n} &
\ds \alpha < \frac{1-\zeta}{2}.
\end{cases}
\end{split}
\end{equation} From the computations by Arita, it is straightforward to check that
\begin{equation}
\label{R-asymp-adj} 
R_{n-2, \zeta n}(\alpha) \approx 
\begin{cases} 
\ds \frac{(1-\zeta^2)^2}{16} R_{n, \zeta n}(\alpha) &\quad\text{if } \alpha \geq (1-\zeta)/2,\\[0.3cm] 
\alpha^2(1-\alpha)^2 R_{n, \zeta n}(\alpha) &\quad\text{otherwise},  
\end{cases}
\end{equation}
and that 
\begin{equation}
\label{S-asymp-adj} 
S_{n-2, \zeta n}(\alpha) \approx 
\begin{cases} 
\ds \frac{(1-\zeta^2)^2}{16} S_{n, \zeta n}(\alpha) &\quad\text{if } \alpha \geq (1-\zeta)/2,\\[0.3cm] 
\alpha^2(1-\alpha)^2 S_{n, \zeta n}(\alpha) &\quad\text{otherwise.}  
\end{cases}
\end{equation}
Moreover, from Stirling's formula, one easily obtains that
\[
\ballot{n + \zeta n + a}{n - \zeta n + b} \approx
\frac{\zeta(1+\zeta)^{-a-1}(1-\zeta)^{-b}}{2^{-a-b-1}\sqrt{\pi (1- \zeta^2) n}} 
\left( \frac{4}{1-\zeta^2} \right)^n \left( \frac{1-\zeta}{1+\zeta} \right)^{\zeta n},
\]
so that
\begin{equation}
\label{ballot1}
\ballot{n + \zeta n -3}{n - \zeta n} \approx
\frac{\zeta(1+\zeta)^2}{4\sqrt{\pi (1- \zeta^2) n}} 
\left( \frac{4}{1-\zeta^2} \right)^n \left( \frac{1-\zeta}{1+\zeta} \right)^{\zeta n},
\end{equation}
and
\begin{equation}
\label{ballot2}
\ballot{n + \zeta n -3}{n - \zeta n - 1} \approx
\frac{\zeta(1+\zeta)^2 (1-\zeta)}{8\sqrt{\pi (1- \zeta^2) n}} 
\left( \frac{4}{1-\zeta^2} \right)^n \left( \frac{1-\zeta}{1+\zeta} \right)^{\zeta n}.
\end{equation}

\subsection{Asymptotics for $Z$}
We are now in a position to compute the asymptotics of $Z_{n,n_0}$ using \eqref{Z*-R}--\eqref{ballot2}. 
These depend on the relative values of $\alpha$ and $\beta$ as follows.

\begin{itemize}

\item For $\alpha, \beta > (1-\zeta)/2$, $\alpha \neq \beta$, 
\begin{multline}
\label{pf-asymp-1}
Z_{n, \zeta n} \approx \frac{\zeta(\zeta+1)^{2}(\alpha(2\alpha_*-\zeta+1)+\alpha_* (\zeta-1)) (\beta (2\beta_*-\zeta+1)+\beta_* (\zeta-1))}{4\alpha_* \beta_* (2\alpha+\zeta-1) (2\beta+\zeta-1)\sqrt{\pi(1-\zeta^2)n}} \\
\times \left( \frac{4}{1-\zeta^2} \right)^n \left( \frac{1-\zeta}{1+\zeta} \right)^{\zeta n}.
\end{multline}

\item If $\alpha = \beta > (1-\zeta)/2$, 
\begin{multline}
\label{pf-asymp-2}
Z_{n, \zeta n} \approx \frac{\zeta(\zeta+1)^{2} (\alpha (2 \alpha_*-\zeta+1)+\alpha_* (\zeta-1)) (\alpha (2 \beta_*-\zeta+1)+\beta_* (\zeta-1))}{4\alpha_* \beta_* (2 \alpha+\zeta-1)^2\sqrt{\pi(1-\zeta^2)n}} \\
\times \left( \frac{4}{1-\zeta^2} \right)^n \left( \frac{1-\zeta}{1+\zeta} \right)^{\zeta n}. 
\end{multline}

\item If $\alpha > \beta = (1-\zeta)/2$, 
\begin{equation}
\label{pf-asymp-3}
Z_{n, \zeta n} \approx \frac{\beta(1-\beta)(1-2\beta)(\alpha_*(\beta-\alpha)-\alpha\beta)}{2\alpha_*\beta_*(\beta-\alpha)} \left(\frac{1}{\beta (1-\beta)} \right)^n \left( \frac{\beta}{1-\beta} \right)^{\zeta n}. 
\end{equation}

\item If $\beta > \alpha = (1-\zeta)/2$, 
\begin{equation}
\label{pf-asymp-4}
Z_{n, \zeta n} \approx \frac{\alpha(1-\alpha)(1-2\alpha)(\beta_*(\alpha-\beta)-\beta\alpha)}{2\alpha_*\beta_*(\alpha-\beta)} \left(\frac{1}{\alpha (1-\alpha)} \right)^n \left( \frac{\alpha}{1-\alpha} \right)^{\zeta n}. 
\end{equation}

\item If $\alpha = \beta = (1-\zeta)/2$, 
\begin{equation}
\label{pf-asymp-5}
Z_{n, \zeta n} \approx \frac{\alpha^2(1-2\alpha)}{\alpha_*\beta_*}\sqrt{\frac{\alpha(1-\alpha)n}{\pi}} 
\left(\frac{1}{\alpha (1-\alpha)} \right)^n \left( \frac{\alpha}{1-\alpha} \right)^{\zeta n}. 
\end{equation}

\item If $\alpha < \beta, \alpha < (1-\zeta)/2$, 
\begin{equation}
\label{pf-asymp-6}
Z_{n, \zeta n} \approx \frac{\alpha(1-\alpha)(1-2\alpha)(\beta_*(\alpha-\beta)-\beta\alpha)}{\alpha_*\beta_*(\alpha-\beta)} \left(\frac{1}{\alpha (1-\alpha)} \right)^n \left( \frac{\alpha}{1-\alpha} \right)^{\zeta n}. 
\end{equation}

\item If $\beta < \alpha, \beta < (1-\zeta)/2$, 
\begin{equation}
\label{pf-asymp-7}
Z_{n, \zeta n} \approx \frac{\beta(1-\beta)(1-2\beta)(\alpha_*(\beta-\alpha)-\alpha\beta)}{\alpha_*\beta_*(\beta-\alpha)} \left(\frac{1}{\beta (1-\beta)} \right)^n \left( \frac{\beta}{1-\beta} \right)^{\zeta n}. 
\end{equation}

\item If $\alpha = \beta < (1-\zeta)/2$, 
\begin{equation}
\label{pf-asymp-8}
Z_{n, \zeta n} \approx \frac{\alpha^2(1-2\alpha)(1-\zeta-2\alpha)n}{\alpha_*\beta_*} 
\left(\frac{1}{\alpha (1-\alpha)} \right)^n \left( \frac{\alpha}{1-\alpha} \right)^{\zeta n}. 
\end{equation}

\end{itemize}

\subsection{Asymptotics for $Z^*$}

Quite similar to the second and last terms of \eqref{pf-dstar}, \eqref{pf-sum-1} can be written as 
\begin{multline}
Z^*_{n, n_0} = \frac{1}{\beta_*} \left( \ballot{n+n_0-3}{n-n_0-1}  +  
R_{n-2,n_0}(\beta) \right) + \frac{\alpha\beta}{\alpha_*\beta_*(\beta-\alpha)}\left(R_{n-2, n_0}(\alpha)-R_{n-2,n_0}(\beta) \right),
\end{multline}
if $\alpha \neq \beta$, and
\begin{equation}
Z^*_{n, n_0} = \frac{1}{\beta_*} \left( \ballot{n+n_0-3}{n-n_0-1}  +  
R_{n-2,n_0}(\beta) \right) + \frac{1}{\alpha_*\beta_*}S_{n-2, n_0}(\beta),
\end{equation}
if $\alpha = \beta$.
The asymptotics of $Z^*_{n, \zeta n}$ can now be computed like the asymptotics of $Z_{n, \zeta n}$. 

\begin{itemize}

\item If $\alpha, \beta > (1-\zeta)/2$, $\alpha \neq \beta$, 
\begin{multline}
\label{pf1-asymp-1}
Z^*_{n, \zeta n} \approx \frac{\alpha(1-\zeta)\zeta(1+\zeta)^2(\beta(2\alpha_*-\zeta+1)+\alpha_* (\zeta-1))}{4\alpha_* \beta_* (2\alpha+\zeta-1) (2\beta+\zeta-1)\sqrt{\pi(1-\zeta^2)n}} \\
\times \left( \frac{4}{1-\zeta^2} \right)^n \left( \frac{1-\zeta}{1+\zeta} \right)^{\zeta n}.
\end{multline}

\item If $\alpha = \beta > (1-\zeta)/2$, 
\begin{multline}
\label{pf1-asymp-2}
Z^*_{n, \zeta n} \approx \frac{\beta(1-\zeta)\zeta(1+\zeta)^2(\beta(2\alpha_*-\zeta+1)+\alpha_*(\zeta-1))}{4\alpha_*\beta_*(2\beta+\zeta-1)^2\sqrt{\pi(1-\zeta^2)n}}\\ 
\times\left( \frac{4}{1-\zeta^2} \right)^n \left( \frac{1-\zeta}{1+\zeta} \right)^{\zeta n}. 
\end{multline}

\item If $\alpha > \beta = (1-\zeta)/2$, 
\begin{equation}
\label{pf1-asymp-3}
Z^*_{n, \zeta n} \approx \frac{\beta(1-\beta)(1-2\beta)(\alpha_*(\beta-\alpha)-\alpha\beta)}{2\alpha_*\beta_*(\beta-\alpha)} \left(\frac{1}{\beta (1-\beta)} \right)^n \left( \frac{\beta}{1-\beta} \right)^{\zeta n}. 
\end{equation}

\item If $\beta > \alpha = (1-\zeta)/2$, 
\begin{equation}
\label{pf1-asymp-4}
Z^*_{n, \zeta n} \approx \frac{\alpha^2\beta(1-\alpha)(1-2\alpha)}{2\alpha_*\beta_*(\beta-\alpha)} 
\left(\frac{1}{\alpha (1-\alpha)} \right)^n \left( \frac{\alpha}{1-\alpha} \right)^{\zeta n}. 
\end{equation}

\item If $\alpha = \beta = (1-\zeta)/2$, 
\begin{equation}
\label{pf1-asymp-5}
Z^*_{n, \zeta n} \approx \frac{\alpha^2(1-2\alpha)}{\alpha_*\beta_*}\sqrt{\frac{\alpha(1-\alpha)n}{\pi}} 
\left(\frac{1}{\alpha (1-\alpha)} \right)^n \left( \frac{\alpha}{1-\alpha} \right)^{\zeta n}.
\end{equation}

\item If $\alpha < \beta, \alpha < (1-\zeta)/2$, 
\begin{equation}
\label{pf1-asymp-6}
Z^*_{n, \zeta n} \approx \frac{\alpha^2\beta(1-\alpha)(1-2\alpha)}{\alpha_*\beta_*(\beta-\alpha)} 
\left(\frac{1}{\alpha (1-\alpha)} \right)^n \left( \frac{\alpha}{1-\alpha} \right)^{\zeta n}. 
\end{equation}

\item If $\beta < \alpha, \beta < (1-\zeta)/2$, 
\begin{equation}
\label{pf1-asymp-7}
Z^*_{n, \zeta n} \approx \frac{\beta(1-\beta)(1-2\beta)(\alpha_*(\beta-\alpha)-\alpha\beta)}{\alpha_*\beta_*(\beta-\alpha)} \left(\frac{1}{\beta (1-\beta)} \right)^n \left( \frac{\beta}{1-\beta} \right)^{\zeta n}. 
\end{equation}

\item If $\alpha = \beta < (1-\zeta)/2$, 
\begin{equation}
\label{pf1-asymp-8}
Z^*_{n, \zeta n} \approx \frac{\alpha^2(1-2\alpha)(1-\zeta-2\alpha)n}{\alpha_*\beta_*} 
\left(\frac{1}{\alpha (1-\alpha)} \right)^n \left( \frac{\alpha}{1-\alpha} \right)^{\zeta n}. 
\end{equation}
\end{itemize}

\subsection{Asymptotics for $Z^0$}
Similar to the second term of \eqref{pf-dstar}, \eqref{pf-sum-2} becomes 
\begin{equation}
Z^0_{n, n_0} = \frac{1}{\alpha_*} \left( \ballot{n+n_0-3}{n-n_0-1}  +  
R_{n-2,n_0}(\alpha) \right) + \ballot{n+n_0-3}{n-n_0}.
\end{equation}
The asymptotics of $Z^0_{n, \zeta n}$ can now also be computed in a similar manner, and 
the results are summarized below. 
\begin{itemize}
\item If $\alpha > (1-\zeta)/2$, 
\begin{multline}
\label{pf2-asymp-1}
Z^0_{n, \zeta n} \approx \frac{\zeta(1+\zeta)^2(\alpha(2\alpha_*+1-\zeta)+\alpha_*(\zeta-1))}{4\alpha_*(2\alpha-1+\zeta)\sqrt{\pi(1-\zeta^2)n}} \left( \frac{4}{1-\zeta^2} \right)^n \left( \frac{1-\zeta}{1+\zeta} \right)^{\zeta n}. 
\end{multline}

\item If $\alpha = (1-\zeta)/2$, 
\begin{equation}
\label{pf2-asymp-2}
Z^0_{n, \zeta n} \approx \frac{\alpha(1-\alpha)(1-2\alpha)}{2\alpha_*}
\left(\frac{1}{\alpha (1-\alpha)} \right)^n \left( \frac{\alpha}{1-\alpha} \right)^{\zeta n}. 
\end{equation}

\item If $\alpha < (1-\zeta)/2$, 
\begin{equation}
\label{pf2-asymp-3}
Z^0_{n, \zeta n} \approx \frac{\alpha(1-\alpha)(1-2\alpha)}{\alpha_*}
\left(\frac{1}{\alpha (1-\alpha)} \right)^n \left( \frac{\alpha}{1-\alpha} \right)^{\zeta n}. 
\end{equation}
\end{itemize}

\end{document}